\documentclass[aps,pra,twocolumn,longbibliography,footinbib]{revtex4-1}
\usepackage{graphicx}
\usepackage{MnSymbol}
\usepackage{color}
\usepackage{bm}  % bold math

\newcommand {\Imag} {\mathrm{Im}}

\newcommand {\eff} {\mathrm{eff}}

\newcommand {\INLN} {Universit\'e C\^ote d'Azur, CNRS, INPHYNI, France}
\begin{document}

\title{Population of collective modes in light scattering by many atoms}

\author{William Guerin}
\email{william.guerin@inphyni.cnrs.fr}
\affiliation{\INLN}
\author{Robin Kaiser}
\affiliation{\INLN}

\date{\today}

\begin{abstract}
The interaction of light with an atomic sample containing a large number of particles gives rise to many collective (or cooperative) effects, such as multiple scattering, superradiance and subradiance, even if the atomic density is low and the incident optical intensity weak (linear optics regime). Tracing over the degrees of freedom of the light field, the system can be well described by an effective atomic Hamiltonian, which contains the light-mediated dipole-dipole interaction between atoms. This long-range interaction is at the origin of the various collective effects, or of collective excitation modes of the system.
Even though an analysis of the eigenvalues and eigenfunctions of these collective modes does allow distinguishing superradiant modes, for instance, from other collective modes, this is not sufficient to understand the dynamics of a driven system, as not all collective modes are significantly populated.
%However, the collective modes alone are not sufficient to understand the dynamics of the system, since all collective modes are not significantly populated.
Here, we study how the excitation parameters, i.e. the driving field, determines the population of the collective modes. We investigate in particular the role of the laser detuning from the atomic transition, and demonstrate a simple relation between the detuning and the steady-state population of the modes. This relation allows understanding several properties of cooperative scattering, such as why superradiance and subradiance become independent of the detuning at large enough detuning without vanishing, and why superradiance, but not subradiance, is suppressed near resonance.

\end{abstract}

%\pacs{32.70.Jz, 42.25.Dd,  42.50.Nn}  % PACS, the Physics and Astronomy Classification Scheme.
% 32.50.+d 	Fluorescence, phosphorescence (including quenching)
% 32.70.Jz 	Line shapes, widths, and shifts
% 32.80.Qk 	Coherent control of atomic interactions with photons
% 42.25.Dd 	Wave propagation in random media
% 42.25.Fx 	Diffraction and scattering
% 42.50.Ct 	Quantum description of interaction of light and matter; related experiments
% 42.50.Gy 	Effects of atomic coherence on propagation, absorption, and amplification of light; electromagnetically induced transparency and absorption
% 42.50.Nn 	Quantum optical phenomena in absorbing, amplifying, dispersive and conducting media; cooperative phenomena in quantum optical systems

\maketitle

\section{Introduction}

Collective effects in light scattering by atomic ensembles are at the focus of intense research, both theoretically and experimentally~\cite{Guerin:JMO}. Recently, the question of light localization in atomic media has been the subject of several studies based on an effective Hamiltonian approach~\cite{Rusek:1996,Rusek:2000,Pinheiro:2004,Skipetrov:2014,Bellando:2014,Skipetrov:2015,Maximo:2015,Skipetrov:2016c}. From a total Hamiltonian describing a system of $N$ atoms with at most one quantum of excitation (one photon), the degrees of freedom of the light field are traced over to get an effective non-Hermitian atomic Hamiltonian $H_{\eff}$. In this approach, the eigenmodes and eigenvalues of $H_{\eff}$ are computed and analyzed. However, these quantities are not direct experimental observables, which makes the interpretation more difficult, in particular because the way the initial excitation entered the system is not specified. In a real experiment, the system is driven or excited by some external field and the outcome of the experiment depends on the parameters of this field.

% \textit{Should we add more detail about $H_{\eff}$, like writing it and the approximations behind it ?}

Another, complementary approach has been used recently in the context of single-photon superradiance~\cite{Scully:2006,Scully:2009,Araujo:2016,Roof:2016}: coupled-dipole equations (CDEs)~\cite{Javanainen:1999,Svidzinsky:2010}. This approach is based on the same effective Hamiltonian, but adding an external driving field is straightforward~\cite{Courteille:2010,Bienaime:2011,Bienaime:2013}. This describes the dynamics of the system in the low-intensity regime of excitation (linear optics) and allows computing experimental observables, such as the emission diagram~\cite{Courteille:2010,Bienaime:2011}, collective line shape and width~\cite{Chomaz:2012,Javanainen:2014,Meir:2014,Bromley:2016,Jennewein:2016,Zhu:2016,Sutherland:2016}, or the temporal dynamics of the scattered light~\cite{Bienaime:2012,Guerin:2016,Araujo:2016,Roof:2016,Skipetrov:2016b}.

The coupled-dipole equations including the external drive read:
\begin{equation}\label{eq.betas}
\dot{\beta}_i = \left( i\Delta-\frac{\Gamma_0}{2} \right)\beta_i -\dfrac{i\Omega(\bm{r})}{2} + \frac{i\Gamma_0}{2} \sum_{i \neq j} V_{ij}(r_{ij})\beta_j \; ,
\end{equation}
where $\beta_i$ is the amplitude of the single-excited-atom state $|i\rangle = |g \cdots e_i \cdots g\rangle$ with $|g\rangle$ ($|e\rangle$) denoting the ground (excited) state, $\Delta$ is the detuning of the driving field from the two-level atomic dipolar transition, $\Omega(\bm{r}) = -dE(\bm{r})/\hbar$ its complex Rabi frequency with $E(\bm{r})$ the driving electric field, $\Gamma_0$ the natural decay rate for a single excited atoms, and $V_{ij}(r_{ij})$ is the dipole-dipole interaction (DDI) between atoms $i$ and $j$, which depends on their separation $r_{ij}$. We will set $\Gamma_0=1$ and drop it in the following.
The first term of Eq.~(\ref{eq.betas}) corresponds to the natural evolution of the dipoles (oscillation and damping), the second one to the driving by the external laser, and the last term corresponds to the DDI interaction.

In the CDEs, the detuning $\Delta$ of the driving field is taken into account, but all collective effects~\cite{Guerin:JMO} -- the trivial ones like the refractive index and the beam attenuation, as well as the non-trivial ones like multiple scattering, super- and sub-radiance -- come from the DDI term, in which the detuning does not directly enter. Since many collective effects obviously depend on the detuning, this can seem puzzling. Moreover, the long-lived modes discussed in the effective Hamiltonian approach~\cite{Skipetrov:2014,Bellando:2014,Skipetrov:2015,Maximo:2015} may be given different interpretations depending on the detuning (e.g., radiation trapping near resonance~\cite{Labeyrie:2003}, or subradiance far from resonance~\cite{Guerin:2016}), although the eigenmodes themselves do not depend on the detuning. % since the detuning appears in the CD equations as a constant shift of the diagonal terms.
Understanding the influence of the detuning is thus crucial to make the link between the CDE and the effective Hamiltonian approach.

In this paper, we study the influence of the detuning on the \emph{populations} of the collective modes, a quantity that has been overlooked so far, except in very few works~\cite{Li:2013,Feng:2014}. In Sec.~\ref{sec.algebra} we derive a simple and intuitive analytical expression relating the steady-state mode populations and the detuning. Although the result [Eq.~(\ref{eq.pk})] is well-known, we show in Sec.~\ref{sec.qualitative} that is has interesting and non-obvious consequences. In particular, it allows us to understand why cooperative effects such as super- and subradiance become independent of the detuning at large detuning and why superradiance vanishes near resonance but not subradiance. Those behaviors are not intuitive and have already been observed experimentally and numerically~\cite{Bienaime:2012,Guerin:2016,Araujo:2016}. We also show that subradiance and radiation trapping~\cite{Labeyrie:2003} can be attributed to collective modes with different eigenvalues, an interpretation supported by the shape of the corresponding eigenmodes.

%Finally, a systematic analysis allows us to put in evidence an empirical scaling law relating ... ???????????

%More deeply, I think this issue is very important to clarify, because most people associate the $1/(kr)^2$ and $1/(kr)^3$ terms of the dipole-dipole interaction to 'cooperative effects', which become negligible at low density, and the long-range $1/kr$ term to 'multiple scattering', which is important when the optical thickness is large. This is somewhat misleading and it's better to call them respectively near-field and far-field terms, but in both cases it's indeed the field radiated by one dipole seen by the other one. So, it is tempting to associate the far-field term to a real 'photon' exchange (in opposition to virtual photon for the near field terms), which sounds very much like 'multiple scattering', which should become negligible if b( )<1.
%On the contrary, our claim is that some cooperative effects remain in the low density limit (with only the 1/kr term), that they are different from the collective effects due to multiple scattering (we claim that radiation trapping is different from subradiance), and that they depend on b0 only, independently of $\Delta$  (with some normalization of course). How is that possible? This is related to the underlying mode structure, which is independent of  $\Delta$, for which speaking of photon is not really appropriate. The problem is indeed purely classical.

\section{Analytical result}\label{sec.algebra}

Let us first write the CDEs [Eq.~\ref{eq.betas}] in a matrix form,
\begin{equation}
\dot{\mathbf{B}} = \mathbf{M}\times \mathbf{B} + \boldsymbol{\Omega} \; ,
\end{equation}
with $\mathbf{B} = [\beta_1, \ldots , \beta_i, \ldots, \beta_N]^\intercal$, $\boldsymbol{\Omega} = -i/2 \times [\Omega(\bm{r_1}), \ldots, \Omega(\bm{r_i}), \ldots , \Omega(\bm{r_N})]^\intercal $, and
\begin{equation}
\mathbf{M} = \begin{bmatrix} -1/2 + i\Delta &  \ldots & V_{1,N} \\ V_{2,1} & \ldots & V_{2,N} \\ \vdots & \ddots & \vdots \\ V_{N,1} & \ldots &  -1/2 + i\Delta \end{bmatrix} \, .
\end{equation}

We note that the detuning $\Delta$ appears as a constant shift of the imaginary part of the diagonal elements of the coupling matrix $\mathbf{M}$. As a consequence, it corresponds to a constant shift of all eigenfrequencies and does not change the eigenvectors. That is the reason why its influence is not discussed in the effective Hamiltonian approach~\cite{Rusek:1996,Rusek:2000,Pinheiro:2004,Skipetrov:2014,Bellando:2014,Skipetrov:2015,Maximo:2015,Skipetrov:2016c}, in which $H_\eff = i \hbar \mathbf{M}(\Delta=0)$ is used, although the correct definition of $H_\eff$ should in principle include the detuning~\cite{Rotter:2009}.

By definition, the eigenvalues $\lambda_k$ and eigenvectors $\mathbf{V}_k$ are such that
$\mathbf{M} = \mathbf{V} \mathbf{D} \mathbf{V}^{-1}$ and $\mathbf{D} = \mathbf{V}^{-1} \mathbf{M} \mathbf{V}$,
where $\mathbf{D} = \mathrm{diag}(\lambda_1 ,\ldots, \lambda_k, \ldots , \lambda_N)$ and $\mathbf{V} = [\mathbf{V}_1, \ldots , \mathbf{V}_k, \ldots , \mathbf{V}_N ]$.

Many experiments~\cite{Labeyrie:2003,Guerin:2016,Araujo:2016} consist in studying the dynamics of the system when it relaxes from the steady state to the ground state after the switch-off of the driving laser. This dynamics is then given by the natural evolution of each mode,
\begin{equation}
\mathbf{B}(t) = \sum_k \alpha_k \mathbf{V_k} e^{\lambda_k t}\; ,
\end{equation}
where the $\alpha_k$ are the complex coefficients of each mode, as given by the initial condition. In the case we consider here, the initial condition corresponds to the steady state reached when the driving laser is on. Let us call this steady state $\mathbf{B_0}$. Obviously,
\begin{equation}\label{eq.B0}
\mathbf{B_0} = -\mathbf{M}^{-1} \boldsymbol{\Omega} = -(\mathbf{V} \mathbf{D} \mathbf{V}^{-1})^{-1} \boldsymbol{\Omega} = -\mathbf{V} \mathbf{D}^{-1} \mathbf{V}^{-1} \boldsymbol{\Omega} \, .
\end{equation}
Let us also project the steady state $\mathbf{B_0}$ on the eigenmodes of the system, we have
\begin{equation}
\mathbf{B_0} = \sum_k \alpha_k \mathbf{V_k} \; ,
\end{equation}
where the coefficients of the decomposition are
\begin{equation}
\boldsymbol{\alpha} = [\alpha_1, \ldots, \alpha_k, \ldots, \alpha_N] = \mathbf{V}^{-1} \mathbf{B_0} \, .
\end{equation}
Using the expression (\ref{eq.B0}) above for $\mathbf{B_0}$, $ \boldsymbol{\alpha} = -\mathbf{D}^{-1} \mathbf{V}^{-1} \boldsymbol{\Omega}$, and we obtain, using the fact that $\mathbf{D}$ is diagonal,
\begin{equation}\label{eq.alpha_k}
\alpha_k = - \frac{(\mathbf{V}^{-1} \boldsymbol{\Omega})_k}{\lambda_k} = - \frac{P_k(\boldsymbol{\Omega})}{\lambda_k},
\end{equation}
where we defined $P_k(\boldsymbol{\Omega}) = (\mathbf{V_k}^\intercal \mathbf{V_k})^{-1}\mathbf{V_k}^\intercal \boldsymbol{\Omega}$ the projection of $\boldsymbol{\Omega}$ onto $\mathbf{V_k}$.
%{\color{blue} I don't know what is the best notation for $(\mathbf{V}^{-1} \boldsymbol{\Omega})_k$, that's also $=(\mathbf{V_k}^\intercal \mathbf{V_k})^{-1}\mathbf{V_k}^\intercal \boldsymbol{\Omega}$, where the first term is the norm and the second the scalar product, but without conjugate...}

This relation is interesting because the weight of each eigenmode in the steady state appears as the product of two factors, one purely ``geometrical'', $\mathbf{V}^{-1} \boldsymbol{\Omega}$, which is the projection of the driving field on the eigenmodes, independent of the detuning, and one purely ``spectral'', the inverse of the corresponding eigenvalue, which does depend on the detuning.
Defining the ``population'' $p_k = |\alpha_k|^2$ of the modes, and noting $\lambda_k = -\Gamma_k/2 + iE_k$, we have
\begin{equation}\label{eq.pk}
p_k = \frac{|P_k(\boldsymbol{\Omega})|^2}{\Gamma_k^2/4 + \left(E_k^0 + \Delta \right)^2} \, ,
\end{equation}
where $E_k^0$ is the eigenfrequency for $\Delta=0$ as in the $H_\eff$ approach. We recover an intuitive result, which describes a Lorentzian coupling efficiency to each mode. This Lorentzian depends on the width of the modes $\Gamma_k$ and is shifted by the detuning $\Delta$.

%It seems to be an obvious result, but, as discussed in the following, it has interesting consequences.

Note that this derivation and the result of Eqs.~(\ref{eq.alpha_k},\ref{eq.pk}) is a simple example of the more general relations that exist between the effective Hamiltonian, its related scattering matrix and decay rates, which are well known in the context of open quantum systems (see the reviews~\cite{Dittes:2000,Okolowicz:2003,Rotter:2009,Kuhl:2012} and, for instance, Eq.~(44) of \cite{Rotter:2009}). The similarities between cooperative scattering and the physics of open quantum systems has started to be discussed only recently~\cite{Fortschritt2012,Rotter:2015}. Here, we just aim at discussing the consequences of this result on the decay of the steady-state after the driving laser is switched off, and in particular the role of the initial detuning of the laser.

\section{Qualitative analysis}\label{sec.qualitative}

To fully understand the consequences of this result, let us turn to some graphical representations, where we plot the eigenvalue distribution of the coupling matrix in the complex plane. The main consequence of Eq.~(\ref{eq.pk}) is that the spectral factor $1/|\lambda_k|^2$ favors the modes located near the origin $E_k=\Gamma_k=0$.

In the following, for simplicity, we will focus on the dilute limit, we thus do not discuss the localization problem~\cite{Rusek:1996,Rusek:2000,Pinheiro:2004,Skipetrov:2014,Bellando:2014,Skipetrov:2015,Maximo:2015,Skipetrov:2016c}, and we discard the near-field terms of the DDI, which are negligible in this limit. Our investigation is thus relevant, for example, to discuss the difference between subradiance~\cite{Guerin:2016} and radiation trapping~\cite{Labeyrie:2003}, or the suppression of superradiance near resonance, as observed in a recent experiment~\cite{Araujo:2016}. In this dilute limit the DDI term is
\begin{equation}
V_{ij}(r_{ij}) = \frac{e^{ikr_{ij}}}{kr_{ij}} \; , \quad r_{ij} = |\bm{r}_i - \bm{r}_j| \; , \label{eq.Vij}
\end{equation}
where $k=2\pi/\lambda$ is the wavevector of the associated atomic transition.

Still for simplicity, we will also take the electric field $E(\bm{r})$ as a plane wave such that
\begin{equation}
\boldsymbol{\Omega} = -\frac{i\Omega}{2} \left[ e^{ikz_1}, \ldots, e^{ikz_i} , \ldots , e^{ikz_N}  \right]^\intercal \, .
\end{equation}
%We will study the influence of the size of the driving field in the next section.

We draw $N$ random positions for the atoms in a spherical Gaussian distribution (rms width $R$) such that the density varies smoothly, thus avoiding sharp edges responsible for internal reflection of light~\cite{Bachelard:2012,Schilder:2016}. We also apply an exclusion volume $k r_{ij} > 3$ to avoid pairs of very close atoms responsible for subradiant and superradiant branches in the complex plane~\cite{Rusek:2000,Skipetrov:2011,Bellando:2014}. Then we diagonalize the coupling matrix $\mathbf{M}$ and compute the weight of the different modes using Eq.~(\ref{eq.pk}).

\begin{figure}[t]
\centering
\includegraphics{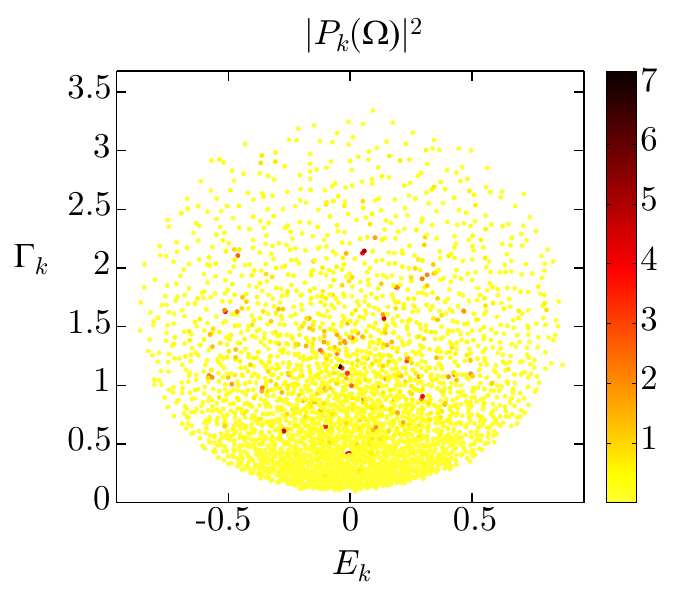}
\caption{Distribution of the eigenvalues $\lambda_k = -\Gamma_k/2 + iE_k$ in the complex plane (for $\Delta=0$) with the geometrical factor $|P_k(\boldsymbol{\Omega})|^2$ represented in the color scale. The parameters of the atomic sample are $N=3000$, $kR \simeq 26.3$, yielding $b_0 \simeq 8.7$ and $\rho_0 k^{-3} \simeq 10^{-2}$.}
\label{fig.geometrical_factor}
\end{figure}

\begin{figure}[t]
\centering
\includegraphics{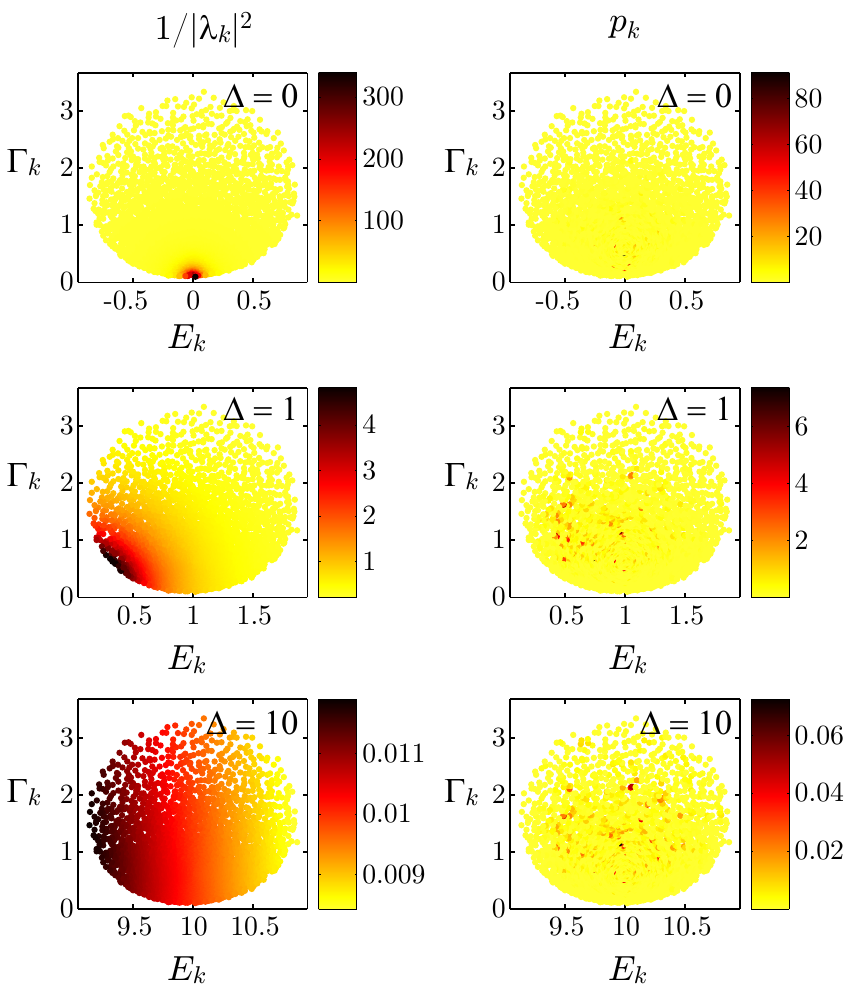}
\caption{Same as in Fig.~\ref{fig.geometrical_factor} but the color scale now shows the spectral factor $1/|\lambda_k|^2$ (left column) and the populations $p_k$ (right column). The different rows are for different detunings, from top to bottom $\Delta = 0, 1, 10$. Note the different color scale for each panel, showing that at large detuning, the spectral factor is almost uniform.}
\label{fig.big_fig}
\end{figure}

\subsection{Influence of the detuning}\label{sec.detuning}

We show in Figs.~\ref{fig.geometrical_factor}-\ref{fig.big_fig} the outcome of such a computation, in which each panel shows the eigenvalue distribution in the complex plane for a single realization of the positions. Similar distributions have been studied before~\cite{Rusek:1996,Rusek:2000,Pinheiro:2004,Skipetrov:2011,Skipetrov:2014,Bellando:2014,Skipetrov:2015,Maximo:2015,Skipetrov:2016c}.
It is known that the eigenvalue distribution spreads from the single-atom limit $\{ E_k=\Delta, \Gamma_k=1 \}$ as the on-resonance optical thickness $b_0 = 2N/(k_0R)^2$ increases,
%~\footnote{For consistency with previous and experimental papers, we define $b_0 = 3N/(k_0 R)^2$, although in the scalar model that we use for the DDI interaction [Eq.~(\ref{eq.Vij})], the actual on-resonance optical thickness is $2N/(k_0 R)^2$.}
first forming a disk of radius $\propto \sqrt{b_0}$ for $b_0 \ll 1$, and then deforms at high $b_0$ with an accumulation of eigenvalues at small $\Gamma_k \ll 1$ and a corresponding spreading at high $\Gamma_k > 1$~\cite{Skipetrov:2011}. This departure from single-atom physics exists at low density and is responsible for many collective effects in light scattering~\cite{Guerin:JMO}.

Here, we also show the geometrical factor $|P_k(\boldsymbol{\Omega})|^2$ (Fig.~\ref{fig.geometrical_factor}), the spectral factor $1/|\lambda_k|^2$ (Fig.~\ref{fig.big_fig}, left column), and the population of the modes $p_k$ (Fig.~\ref{fig.big_fig}, right column) encoded in the color scale. The different rows of Fig.~\ref{fig.big_fig} are for different detunings, on resonance $\Delta=0$ (first row), slightly detuned $\Delta = 1$ (second row) and far detuned $\Delta = 10$ (third row). The geometrical factor (Fig.~\ref{fig.geometrical_factor}) does not depend on the detuning. Here we have chosen a moderately large on-resonance optical thickness $b_0 \simeq 9$ and a low density $\rho_0 k^{-3} \simeq 10^{-2}$ ($\rho_0$ is the peak atomic density). Since the problem is linear, the value of $\Omega$ can be chosen at will and we have taken $\Omega = 2/\sqrt{N}$ such that $\boldsymbol{\Omega}$ is normalized to unity.

From these figures, several relevant observations can be made:
(1) Only a few modes, mainly selected by the geometrical factor, have a nonnegligible population and thus contribute to the dynamics of the system. Studying the whole eigenvalue distribution is thus not directly relevant to the experiment. In particular, the extreme modes, for example the most superradiant ones, whose eigenvalues lie on the border of the distribution, are not significantly populated.
%This was to be expected, since the plane wave illumination does not respect the spherical symmetry of the system. {\color{blue} To be checked: is the most superradiant mode symmetric?}
(2) The geometrical factor favors the short-lived modes, i.e., the superradiant modes ($\Gamma_k > 1$). This was expected from the idea that superradiant modes are more coupled to the environment than subradiant modes.
% In this approach, it appears as mainly a geometrical effect (shape and symmetry of the modes), and not a spectral effect (width of the modes). Bof... ça montre que les deux sont liés.
(3) Far from resonance, the spectral factor only induces an overall decrease of the populations and has a negligible effect on the mode selection.
(4) It is very hard, if not impossible, to select any specific mode by tuning the driving field frequency. For the $\Delta=1$ case, for example, one might expect to selectively populate modes on the left border of the distribution, but that is actually not the case. The geometrical factor dominates over the spectral factor. Other strategies, based for instance on the spatial shaping of the driving field, are needed to selectively populate targeted modes~\cite{Scully:2015}.
(5) The spectral factor has an important effect only near resonance. It strongly favors the long-lived modes and thus decreases considerably the relative population of the superradiant modes. This demonstrates that superradiance disappears near resonance, as observed in previous experiments and numerical simulations~\cite{Araujo:2016,Note1}. 
%\footnote{This discussion does not apply to experiments using a pulsed excitation, as in Ref.~\cite{Roof:2016}, since we are dealing with the steady state obtained with a continuous monochromatic excitation.}.

\begin{figure}[t]
\centering
\includegraphics{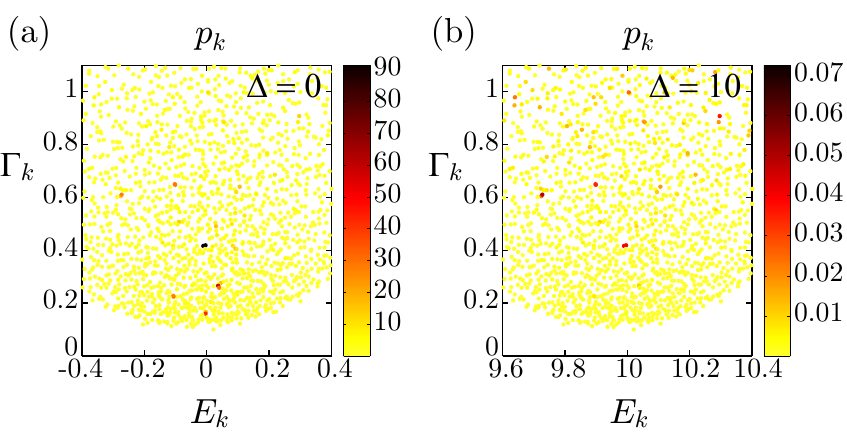}
\caption{Close-up on the populations of the long-lived modes, at resonance (a) and far from resonance (b). Same parameters as in Fig.~\ref{fig.geometrical_factor}.}
\label{fig.long-lived}
\end{figure}

A closer look on the populations of the long-lived modes is shown in Fig.~\ref{fig.long-lived} for the resonant (a) and far-detuned (b) cases. Even on resonance, only a few modes are strongly populated, showing that the geometrical factor still plays an important role. At large detuning, the long-lived modes that are populated are responsible for subradiance. These modes are still populated (and even more) near resonance, showing that the relative weight of subradiance increases near resonance, as seen experimentally~\cite{Guerin:2016}.
Moreover, in addition to the modes strongly populated at large detuning, additional eigenmodes acquire noticeable population near resonance, with even longer lifetimes.
These modes are responsible for radiation trapping due to multiple scattering~\cite{Labeyrie:2003}. This interpretation is validated by an analysis of the spatial properties of the modes, summarized in Fig.~\ref{fig.spatial} and detailed in the next section.

At large detuning, the effect of the spectral factor on the relative population of the eigenmodes is negligible, and completely vanishes for $\Delta \rightarrow \infty$, such that the relative populations are only given by the geometrical factor $|P_k(\boldsymbol{\Omega})|^2$, which actually means that the steady state $\mathbf{B_0}$ is proportional to $\boldsymbol{\Omega}$. This is precisely the ``timed-Dicke'' (TD) state approximation, introduced for single-photon excitation by Scully \textit{et al.}~\cite{Scully:2006} and further developed for continuous driving in Refs.~\cite{Courteille:2010,Bienaime:2011,Bienaime:2013}. Although this state is mainly superradiant, even though not the largest $\Gamma_k$ of the distribution, it also contains subradiant components [Fig.~\ref{fig.long-lived}(b)], as recently observed experimentally~\cite{Guerin:2016}.
In other words, using a large detuning makes the driving field to couple weakly, but \emph{equally}, to all modes having a good geometrical overlap with the driving field, and it thus reveals a part of the underlying mode structure, which is independent of the detuning.
The consequence is that the collective dynamics after switching off the driving field is still cooperative at very large detuning, with superradiant and subradiant decay rates becoming asymptotically independent of the detuning.

\subsection{Spatial properties of the modes} \label{sec.spatial}

It is also interesting to study the spatial properties of the collective modes in order to identify their physical meaning.

\begin{figure}[t]
\centering
\includegraphics{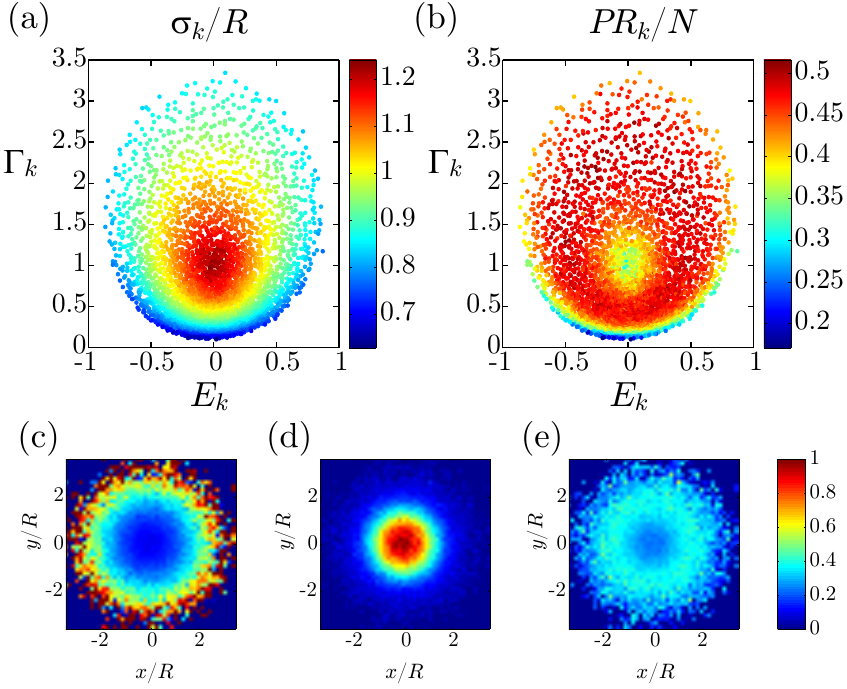}
\caption{Spatial properties of the collective modes. Same parameters are in Fig.~\ref{fig.geometrical_factor}, but the color scale now shows: (a) the rms size $\sigma_k$ of the modes (normalized by the sample size $R$); (b) the participation ratio $PR_k$ (normalized by the atom number $N$). We show in panels (c,d,e) the average over 120 realizations of the excitation probability (mode intensity divided by atomic density) for atoms located in a slice $|z|<R/5$, for each kind of modes, selected by the following conditions: (c) $0.9 < \Gamma_k < 1.1$ and $-0.1 < E_k < 0.1$; (d) $\Gamma_k < 0.2$; (e) $PR_k > 0.45$.}
\label{fig.spatial}
\end{figure}

Two quantities are useful to characterize the mode spatial properties, the rms size of the modes $\sigma_k$ and
the participation ratio (PR), defined as
\begin{equation}
PR_k = \frac{\sum_i |V_{ki}|^2} { \sum_i |V_{ki}|^4 } \; ,
\end{equation}
which indicates the number of atoms participating significantly to the mode~\cite{Skipetrov:2014,Biella:2013}. We represent in Fig.~\ref{fig.spatial}(a,b) these two quantities in the color scale of the eigenvalue distribution. We observe three distinctive areas.
(i) Near the single-atom-physics case $\{ E_k=0, \Gamma_k=1 \}$, the modes have a larger rms size than the Gaussian atomic sample, $\sigma_k > R$, and a local minimum of the PR. This denotes modes delocalized at the boundary of the sample. Physically, this situation corresponds to single-scattering (or low order scattering) on the edges of the sample, as confirmed by the profile shown in Fig.~\ref{fig.spatial}(c).
(ii) For the longest-lived modes ($\Gamma_k \ll 1$, at the border of the distribution), the size and the PR are both small, which means that the modes are not very extended. As seen in Fig.~\ref{fig.spatial}(d), they are located at the center of the sample. We attribute this behavior to diffusive modes due to multiple scattering.
(iii) In the rest of the complex plane (most modes), the modes have approximately the same size as the sample ($\sigma_k \sim R$) and a maximum PR (around $N/2$). They correspond to collective and extended modes, with almost uniform excitation probability across the sample Fig.~\ref{fig.spatial}(d). These modes can exhibit superradiant ($\Gamma_k >1$) or subradiant ($\Gamma_k<1$ or $\Gamma_k \ll 1$ ) behavior.

This analysis validates the interpretation given above on the different nature of the long-lived modes that are populated near resonance (diffusive modes responsible for radiation trapping~\cite{Labeyrie:2003}) compared to those excited far from resonance (subradiant modes).

\section{Numerical study}

\begin{figure*}
\centering
\includegraphics{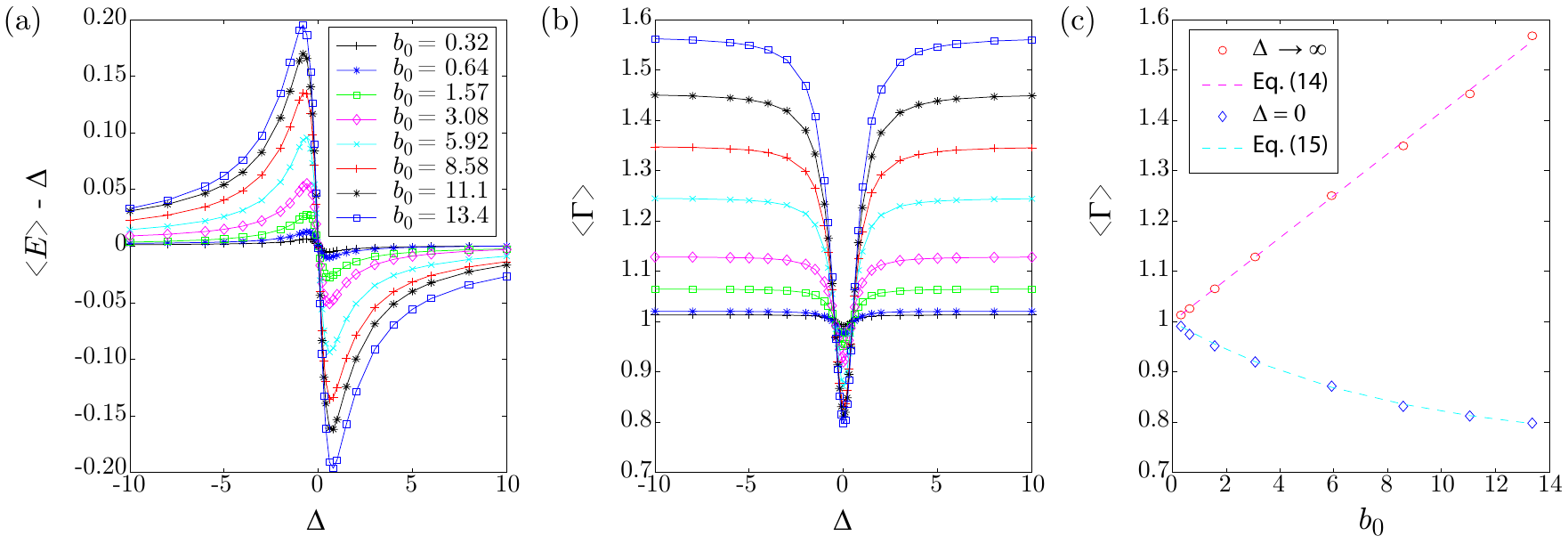}
\caption{Study of the weighted averages $\langle E \rangle$ and $\langle \Gamma \rangle$ [Eq.~(\ref{eq.averaging})] as a function of the detuning $\Delta$ and the on-resonance optical thickness $b_0$. (a) The frequency shift $\langle E \rangle - \Delta$ is plotted as a function of $\Delta$ for different $b_0$. (b) $\langle \Gamma \rangle$ is plotted as a function of $\Delta$ for different $b_0$. (c) $\langle \Gamma \rangle $ is plotted as a function of $b_0$ for the two extreme cases, $\Delta \rightarrow \infty$ (red circles) and $\Delta=0$ (blue diamonds). The dashed lines correspond to the empirical relations given in Eq.~(\ref{eq.Gamma_inf}) and Eq.~(\ref{eq.Gamma_0}), respectively.}\label{fig.Gamma_mean}
\end{figure*}

Many statistical quantities can in principle be computed and studied from the eigenvalue distribution~\cite{Rusek:1996,Rusek:2000,Pinheiro:2004,Skipetrov:2014,Bellando:2014,Skipetrov:2015,Maximo:2015,Skipetrov:2016c,Skipetrov:2011}.
Here we will focus on quantities that uses the information contained in the populations $p_k$. These quantities, not studied before, are thus not only related to the properties of the effective Hamiltonian, but also to the way the system is excited. In particular, they will depend on the detuning $\Delta$. They are thus less universal, but they are more related to experimental observables. One can thus expect to recover qualitative behaviors similar to what has been observed in experiments or in numerical simulations of the CDEs.

\subsection{Behavior of the weighted averages}\label{sec.averages}

Let us now turn to a systematic analysis of the weighted averages
of the eigenfrequencies (or eigenenergies) and decay rates
(or linewidth), defined as:
\begin{equation}\label{eq.averaging}
\langle E \rangle = \frac{\sum_k p_k E_k}{\sum_k p_k } \quad \mathrm{and} \quad \langle \Gamma \rangle = \frac{\sum_k p_k \Gamma_k}{\sum_k p_k } \, .
\end{equation}
We show in Fig.~\ref{fig.Gamma_mean} a systematic study of these quantities as a function of the on-resonance optical thickness $b_0$ and on the detuning $\Delta$. For each $b_0$, 120 realizations of the disorder configuration have been used.

We observe that the average eigenenergy is slightly shifted from $\Delta$ and the shift $\langle E \rangle - \Delta$  displays a dispersion-like behavior, which becomes higher and broader as the optical thickness increases. On the contrary, the average decay rate $\langle \Gamma \rangle$ exhibits a negative resonance-like structure, which is also more important at larger $b_0$. These behaviors are due to the spectral factor and can be qualitatively understood as follows.

First, on resonance ($\Delta=0$), positive and negative values of $E_k$ compensate so that $\langle E \rangle = 0$ after averaging over the disorder configurations (we remain here in the dilute limit such that the cooperative Lamb shift is negligible~\cite{Javanainen:2014,Meir:2014,Bromley:2016,Jennewein:2016,Zhu:2016,Friedberg:1973,Scully:2009b,Rohlsberger:2010,Keaveney:2012,Manassah:2012,Jenkins:2016,Javanainen:2016}). The same applies at very large detuning, for which the spectral factor plays a negligible role on the relative populations. At intermediate detuning, the spectral factor favors one side of the eigenvalue distribution, such that $\langle E \rangle$ departs from $\Delta$ to get slightly closer to zero [Fig.~\ref{fig.big_fig}]. The difference $\langle E \rangle - \Delta$ has thus an opposite sign from $\Delta$, which produces a dispersion behavior for the frequency shift, similar to an effective refractive index. This effect is more important as the eigenvalue distribution spreads for increasing $b_0$.
 %{\color{blue} This frequency shift might be visible in the excitation dynamics of the system [J. Schachenmayer, private communication], to be confirmed.}

Similarly for $\langle \Gamma \rangle$, at large detuning, the geometrical factor dominates and we have seen previously that it favors the superradiant modes, such that $\langle \Gamma \rangle > 1$, and superradiance is stronger as $b_0$ increases. We can actually compute $\langle \Gamma \rangle$ in the $\Delta \rightarrow \infty$ limit (TD approximation) by replacing the populations by the geometrical factor in the averaging Eq.~(\ref{eq.averaging}). We clearly observe in Fig.~\ref{fig.Gamma_mean}(c) a linear scaling with $b_0$:
\begin{equation}\label{eq.Gamma_inf}
\langle \Gamma \rangle_{\Delta \rightarrow \infty} =  1 + \frac{b_0}{24} \, .
\end{equation}
We note that a similar linear scaling is expected for the superradiant decay rate of the TD state~\cite{Mazets:2007,Svidzinsky:2008,Svidzinsky:2008b,Courteille:2010,Friedberg:2010,Prasad:2010,Araujo:2016,Roof:2016}.

On resonance, however, the spectral factor favors the long-lived modes [Fig.~\ref{fig.big_fig}] and thus $\langle \Gamma \rangle$ drops to values smaller than unity. Interestingly, we have found [Fig.~\ref{fig.Gamma_mean}(c)] that the data follow very closely the empirical relationship
\begin{equation}\label{eq.Gamma_0}
\langle \Gamma \rangle_{\Delta=0} =  1- \frac{1}{4}\left(1 - e^{-b_0/8} \right) \, ,
\end{equation}
which denotes an exponential decrease of $\langle \Gamma \rangle$ with the optical thickness, with a saturation effect. This is consistent with the idea that attenuation and multiple scattering suppress superradiance, as observed in ref.~\cite{Araujo:2016} and in Fig.~\ref{fig.big_fig}, but the plane wave illumination insures that there is always a large proportion of single scattering at the borders of the atomic cloud, and thus a large fraction of modes with $\Gamma_k \sim 1$, such that $\langle \Gamma \rangle_{\Delta=0}$ does not decrease to zero as $b_0$ increases.

\begin{figure}
\centering
\includegraphics{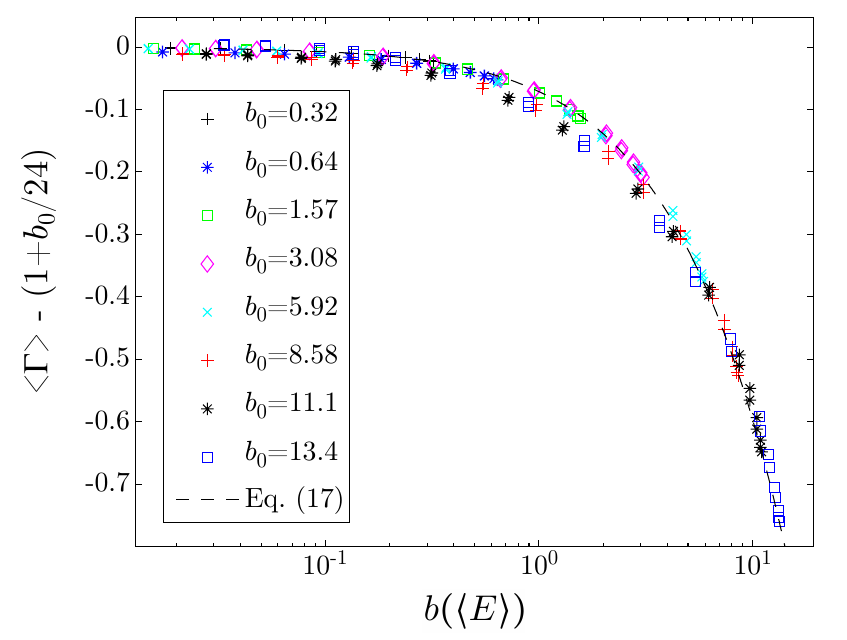}
\caption{Same numerical data as in Fig.~\ref{fig.Gamma_mean}(a,b) but $\langle \Gamma \rangle - (1+b_0/4)$ is plotted as a function of $b(\langle E \rangle^2) = b_0/(1+4\langle E \rangle^2)$ (log scale). All point collapse on a single curve (dashed line), well described by Eq.~(\ref{eq.empirical1}).}
\label{fig.scaling}
\end{figure}

Given the relatively simple behaviors of $\langle \Gamma \rangle$ and $\langle E \rangle$, one can wonder whether a more general relationship between $\langle \Gamma \rangle$, $\langle E \rangle$ and $b_0$ can be found. The limit cases [Eqs.~(\ref{eq.Gamma_inf},\ref{eq.Gamma_0})] suggest a route to a more refined scaling law. Plotting $\langle \Gamma \rangle - \left( 1+ b_0/24 \right)$ as a function of the detuning $\Delta$ exhibits a Lorentzian absorption profile whose depth and width depends on $b_0$. It is thus natural to plot it as a function of $b(\Delta) = b_0/(1+4\Delta^2)$. The points then tend to collapse on a single curve, but with significant deviations. In fact, $b(\Delta)$ would be the actual optical thickness at the laser detuning without cooperativity. To take into account the spreading of the eigenvalue distribution, it makes sense to replace $b(\Delta)$ by
\begin{equation}
b(\langle E \rangle) = \frac{b_0}{1+4\langle E \rangle^2} \, .
\end{equation}
In that case, all data points collapse almost perfectly on a single curve (Fig.~\ref{fig.scaling}). This curve is well described by
\begin{equation}\label{eq.empirical1}
\langle \Gamma \rangle - \left( 1+ \frac{b_0}{24} \right) \simeq  -\frac{b(\langle E \rangle)}{24} - \frac{1}{4}\left(1 - e^{-b(\langle E \rangle)/8} \right) \, ,
\end{equation}
or, in a more compact form, defining $b \equiv b(\langle E \rangle)$,
\begin{equation}\label{eq.empirical}
\langle \Gamma \rangle \simeq 1 + \frac{1}{4} \left[ \frac{b_0 - b}{6} - \left(1 - e^{-b/8} \right) \right] \, ,
\end{equation}
%or, in a more explicit form,
%\begin{equation}
%\langle \Gamma \rangle \simeq 1 + \frac{1}{4} \left\{ \frac{2}{3} \frac{\langle E \rangle^2}{1+4\langle E \rangle^2} b_0 - \left[ 1 - \exp \left( -\frac{b/8}{1+4\langle %E\rangle^2}\right) \right] \right\} \, ,
%\end{equation}
which contains the previous limiting cases.
The quality of the collapse on such a universal curve as seen in Fig.~\ref{fig.scaling} and expressed by Eq.(\ref{eq.empirical}) suggests that it should be possible to obtain analytical results describing the observed behaviors.

\subsection{Linewidth distribution}

%By choosing a threshold on the normalized population, for example $10^{-4}$, one can look at the minimum $\Gamma$.
%By selecting the long-lived modes that are populated in a nonnegligible way (using some criterion), I would like to recover the $\Gamma \sim 1/b_0$ scaling far from resonance (subradiance), and a beginning of $\Gamma \sim 1/b_0^2$ scaling at resonance (radiation trapping). That would be very nice!

% Taking averages does not give anything relevant.\\

Another interesting quantity is the distribution of the linewidths, $P(\Gamma_k)$. This distribution has been studied in several papers~\cite{Kottos:2002,Pinheiro:2004,Weiss:2006,Bellando:2014}, but here we include $p_k$ as a weighting factor of the $\Gamma_k$'s in the distribution $P(\Gamma_k)$, still computed over 120 realizations of the positions.

\begin{figure}[t]
\centering
\includegraphics{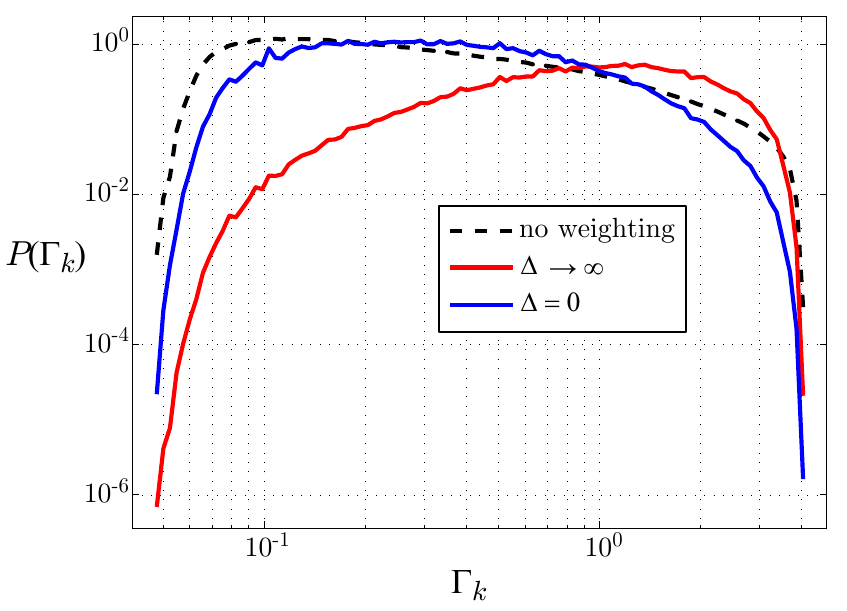}
\caption{Linewidth distribution $P(\Gamma_k)$ without weighting (dashed black line) and with weighting corresponding to the populations $p_k$ computed far from resonance ($\Delta \rightarrow \infty$, red solid line) or at resonance ($\Delta=0$, blue solid line). The parameters of the atomic sample are $N=5000$, $kR \simeq 27.3$ yielding $b_0 \simeq 13.4$, $\rho_0 k^{-3} \simeq 1.55 \times 10^{-2}$, and we used 120 realizations.}
\label{fig.P_Gammas}
\end{figure}

We show in Fig.~\ref{fig.P_Gammas} the comparison of the distribution $P(\Gamma_k)$ without weighting (and thus independent of $\Delta$), with the ones computed with weighting corresponding to $\Delta \rightarrow \infty$ and $\Delta=0$ (and excitation by a plane wave, as previously). As expected from the previous discussions, taking into account the weighting due to the population $p_k$ increases the probability density of the short-lived (superradiant) modes far from resonance, and of the long-lived modes near resonance.
We note that this distribution $P(\Gamma_k)$ has a strong dependance on the detuning.
This suggests that a characterization of the transport properties of light through resonant two level systems needs to go beyond a mere eigenvalues analysis~\cite{Pinheiro:2004}, since the transport properties obviously depend on the detuning. In general, the distribution $P(\Gamma_k)$ depends on the way the system is coupled to the environment~\cite{Dittes:2000,Weiss:2006}.

We also note that the linewidth distribution does not allow us to extract any specific value for the long-lived modes effectively populated by an external drive, even when taking into account the weighting function of Eq.~(\ref{eq.pk}).
Therefore, we cannot recover the scaling laws that can be observed in experiments on subradiance~\cite{Guerin:2016} or radiation trapping \cite{Labeyrie:2003}, indicating that the approach presented in this paper is not sufficient to recover experimentally-observed scaling laws.

\subsection{Discussion}

%Attention, these are not direct observables... Nonorthogonality of the modes, etc. ...

%{\color{blue} PARTIE A REDIGER.}

%The long-lived modes can be experimentally ``selected'' by observing the late-time decay of the system. This is how radiation trapping or subradiance in cold atomic gases have been studied. In the diffusive regime, i.e. for large optical thickness $b(\Delta) \gg 1$, the characteristic decay time of radiation trapping is $\tau_\RT = \Gamma_\RT^{-1} \propto b(\Delta)^2$~\cite{Labeyrie:2003}. At large detuning and large $b_0$, subradiance dominates and a scaling $\tau_\sub = \Gamma_\sub^{-1} \propto b_0$ has been observed experimentally and numerically~\cite{Guerin:2016}. Here, for example at large detuning, the distribution $P(\Gamma_k)$ is smooth and monotonous for $\Gamma_k<1$, and thus does not exhibit any characteristic value.

We attribute this limitation to the fact that the quantities studied in this paper (such as the $\Gamma_k$'s) are not direct experimental observables. Indeed, when measuring the light escaping from the atomic sample, for example the total scattered power
\begin{equation}
\begin{split}
P \propto -\frac{d}{dt} \sum_{i=1}^N \left| \beta_i(t) \right|^2 & = -\frac{d}{dt} \| \mathbf{B}(t) \| \\
& = -\frac{d}{dt} \left\| \sum_k \alpha_k \mathbf{V_k} e^{\lambda_k t} \right\| \; ,
\end{split}
\end{equation}
the nonorthogonality of the eigenmodes $\mathbf{V_k}$ (due to the non-Hermicity of the effective Hamiltonian) are at the origin of oscillating terms~\cite{Okolowicz:2003}, which may change the dynamics of the decay, even after configuration averaging.
It is thus not surprising to find a quantitative difference in the decay rates. For example, the linear scaling of $\langle \Gamma \rangle$ with $b_0$ obtained at Eq.~(\ref{eq.Gamma_inf}) does not have the same slope as what has been found by studying the decay of the scattered light in the coupled-dipole model~\cite{Araujo:2016}.
Obtaining analytical results on experimental observables remains thus an open problem.

%That's why another approach is sometimes used, which is to study the spectrum of the Hermitian operator $\Imag(H_\eff)$~\cite{Akkermans:2008}. In that case, the eigenvalues are directly related to the probability of returning in the ground state, meaning that the light escaped the sample~\cite{Ernst:1968,Ressayre:1977}.
%Is there any conditions ? Is it only for no initial coherence (superfluorescence case) ??? The papers \cite{Ernst:1968,Ressayre:1977} are for superfluorescence. If yes, remove this discussion, since it's not relevant here.

In the framework of the effective Hamiltonian approach, one can also study the spectrum of the Hermitian operator $\Imag(H_\eff)$~\cite{Akkermans:2008}. In that case, the eigenvalues are directly related to the probability of returning in the ground state, and thus correspond to the light escape rates from the sample~\cite{Ernst:1968,Ressayre:1977}. However, this is only valid when the initial state contains one excitation but no coherence, and thus does not apply to driven systems.

% peut-être se limiter à dire qu'il reste des question ouvertes, comme p.ex. le rôle de l'état initial (avec ou sans cohérence) distinguant p.ex. superfluo et superradiance à un photon (et citer [AGK] et JvZanthier: PRL 113, 263606 (2014)]

\section{Conclusion}

Several properties of cooperative scattering, such as the enhancement of subradiance and suppression of superradiance near resonance~\cite{Guerin:2016,Araujo:2016}, and the very existence of cooperative decay at very large detuning, are highly nonintuitive. We have shown in this paper that they are consequences of the simple analytical relationship that exists between the population of the collective modes of the effective Hamiltonian and the detuning of the driving field [Eq.~(\ref{eq.pk})]. We have also put in evidence an empirical scaling law on the weighted averages of the eigenvalues of the effective Hamiltonian, which suggests the possibility of further analytical results.

In general, statistical properties of the eigenvalues of the effective Hamiltonian can be efficiently studied using Random Matrix Theory~\cite{Skipetrov:2011}, whereas very few analytical results have been obtained from the coupled-dipole equation model~\cite{Svidzinsky:2008b,Svidzinsky:2010}. The populations of the collective modes and their dependence with the parameters of the driving field is an important ingredient bridging the gap between the two approaches.

We also briefly discussed the spatial properties of the modes and showed that we can distinguish two kinds of long-lived modes that can be associated to radiation trapping and subradiance. Extending this analysis to the high-density case could be useful to better understand the transition to Anderson localization~\cite{Skipetrov:2014,Bellando:2014}.

% parler aussi de RT, des la forme des modes, de la mise en évidence d'une loi d'échelle empirique (eq 18).

% Conclure que le role des interferences dans RT peut être intéressant et év. permettre de comprendre le passage entre subrad et Anderson...

%We have understood why cooperative effects becomes independent of the detuning at large detuning (it was not obvious, it could have disappeared) and made the link with the Timed-Dicke state used in previous papers~\cite{Scully:2006,Courteille:2010,Bienaime:2013}.
%We have also understood why superradiance is largely suppressed near resonance, as observed in a recent experiment~\cite{Araujo:2016}, but not subradiance, which is, on the contrary, enhanced near resonance, as also observed in a recent experiment~\cite{Guerin:2016}. In that case however, the difference with radiation trapping~\cite{Labeyrie:2003} is not so clear and is the subject of our present experimental study.

%The problem with all this is that the collective modes of the effective Hamiltonian (or the coupling matrix) are not directly related to the time constant for light escaping the sample, because the eigenmodes are not orthogonal...

%So, in the future, we need to extend this kind of approach to the collective modes of the imaginary part of the effective Hamiltonian, which describe better the light escape rate~\cite{Akkermans:2008}.

\begin{acknowledgments}

We acknowledge financial support from the ANR (Agence National pour la Recherche, project LOVE, No. ANR-14-CE26-0032).

\end{acknowledgments}

%\bibliographystyle{D:/RECHERCHE/MesPublis/MaBiblio/prsty_no_etal}
%\bibliography{D:/RECHERCHE/MesPublis/MaBiblio/AllMyBiblio}  % default style is apsrev4-1.bst, use option longbibliography (at the beginning) to have the titles

%% Bibliography with \bibitem

\end{document}